\documentclass[twocolumn,tighten,times]{aastex62}
\usepackage{natbib}
\usepackage[normalem]{ulem}
\received{? ?, ?}
\revised{July 27, 2021}
\accepted{August 5, 2021}

\shorttitle{Magnetic Field Structure over a Plage Region}
\shortauthors{Anan et al.}

\begin{document}

\title{Measurements of Photospheric and Chromospheric Magnetic Field Structures \\ Associated with Chromospheric Heating over a Solar Plage Region}

\correspondingauthor{Tetsu Anan}
\email{tanan@nso.edu}

\author[0000-0001-6824-1108]{Tetsu Anan}
\affil{National Solar Observatory, 22 Ohi`a Ku, Makawao, Hawaii, 96768, USA}

\author[0000-0002-7451-9804]{Thomas A. Schad}
\affil{National Solar Observatory, 22 Ohi`a Ku, Makawao, Hawaii, 96768, USA}

\author{Reizaburo Kitai}
\affil{Ritsumeikan University, 56-1, Toji-in Kitamachi, Kita-ku, Kyoto, 603-8577, Japan}

\author[0000-0002-6003-4646]{Gabriel I. Dima}
\affil{High Altitude Observatory, 3090 Center Green Drive, Boulder, Colorado, 80301, USA}

\author[0000-0001-5459-2628]{Sarah A. Jaeggli}
\affil{National Solar Observatory, 22 Ohi`a Ku, Makawao, Hawaii, 96768, USA}

\author[0000-0002-8259-8303]{Lucas A. Tarr}
\affil{National Solar Observatory, 22 Ohi`a Ku, Makawao, Hawaii, 96768, USA}

\author[0000-0002-6210-9648]{Manuel Collados}
\affil{Instituto de Astrof\'isica de Canarias, 38205 La Laguna, Tenerife, Spain}
\affil{Departamento de Astrof\'isica, Universidad de La Laguna, 38205 La Laguna, Tenerife, Spain}

\author{Carlos Dominguez-Tagle}
\affil{Instituto de Astrof\'isica de Canarias, 38205 La Laguna, Tenerife, Spain}
\affil{Departamento de Astrof\'isica, Universidad de La Laguna, 38205 La Laguna, Tenerife, Spain}

\author[0000-0002-7791-3241]{Lucia Kleint}
\affil{University of Geneva, 7, route de Drize, 1227 Carouge, Switzerland}



\begin{abstract}

 In order to investigate the relation between magnetic structures and the signatures of heating in plage regions, 
we observed a plage region with the \ion{He}{1} 1083.0 nm and \ion{Si}{1} 1082.7 nm lines on 2018 October 3 using the integral field unit mode of the GREGOR Infrared Spectrograph (GRIS) installed at the GREGOR telescope.
During the GRIS observation, the Interface Region Imaging Spectrograph (IRIS) obtained spectra of the ultraviolet \ion{Mg}{2} doublet emitted from the same region.
In the periphery of the plage region, within the limited field of view seen by GRIS, we find that the \ion{Mg}{2} radiative flux increases with the magnetic field in the chromosphere with a factor of proportionality of $2.38 \times 10^4 {\rm \, erg \, cm^{-2} \, s^{-1} \, G^{-1} }$.
The positive correlation implies that magnetic flux tubes can be heated by Alfv\'en wave turbulence or by collisions between ions and neutral atoms relating to Alfv\'en waves. 
Within the plage region itself, the radiative flux was large between patches of strong magnetic field strength in the photosphere, or at the edges of magnetic patches.
On the other hand, we do not find any significant spatial correlation between the enhanced radiative flux and the chromospheric magnetic field strength or the electric current.
In addition to the Alfv\'en wave turbulence or collisions between ions and neutral atoms relating to Alfv\'en waves, other heating mechanisms related to magnetic field perturbations produced by interactions of magnetic flux tubes could be at work in the plage chromosphere.
\end{abstract}

\keywords{ \href{http://astrothesaurus.org/uat/1987}{Solar chromospheric heating (1987)}; \href{http://astrothesaurus.org/uat/1503}{Solar magnetic fields (1503)}; \href{http://astrothesaurus.org/uat/1533}{Solar ultraviolet emission(1533)}; \href{http://astrothesaurus.org/uat/1973}{Spectropolarimetry (1973)} }


\section{Introduction} \label{sec:intro}


Since the temperature in the solar chromosphere and corona is higher than that expected for a state of radiative equilibrium, some mechanical heating is required to maintain the atmospheric temperature \citep{athay66, withbroe77, vernazza81, anderson89}.
The strongest quasi-steady heating occurs in the active chromosphere and in particular within plage regions \citep{lockyer1869, shine74, fontenla91}, which emit brightly in chromospheric lines such as H${\rm \alpha}$ (\citeauthor{chintzoglou21} \citeyear{chintzoglou21} further discuss the definition of plage).
To balance the large energy loss due to radiative cooling from the spectral lines, a heating energy flux of $\sim 6 \times 10^7 {\rm \,erg\, cm^{-2}\, s^{-1}}$ is required to maintain the temperature in these regions \citep{avrett81}.
\citet{howard59} and \citet{leighton59} found that spatial patterns of strong magnetic field in the photosphere correspond to regions with enhanced line emission.
In addition, the Poynting flux in the photosphere is large enough to heat the chromosphere \citep{welsch15}.
Hence, chromospheric heating in plage regions should be related to the magnetic field; however, the details of the energy transport and dissipation are less clear \citep[see, e.g.,][]{carlsson19}.

One goal of this work is to observationally discriminate between the various heating mechanisms at work in plage.
This, however, is a difficult task as both the dynamical and dissipation length scales are unresolved.
High-resolution Interface Region Imaging Spectrograph \citep[IRIS, ][]{pontieu14} observations of the \ion{Mg}{2} h \& k spectral profiles within plage are only currently reproduced using models with high micro-turbulence \citep{Carlsson15, Pontieu15, Cruz16, Santos20}.
This implies that shocks, torsional motions, or Alfv${\rm \acute{e}}$n wave turbulence can occur along the line of sight, or on a spatial scale smaller than 250 km, and may play significant roles in the energy dissipation.
As a result, many potential interrelated processes have been identified, which we will now review in order to identify possible observational discriminants.

Table \ref{table.mechanism1} lists the various heating mechanisms proposed for the chromosphere above regions that consist of magnetic flux tubes with strong magnetic field, i.e., plage or network.
In addition, the table includes how each mechanism's heating rate is expected to scale with the magnetic field.

\begin{deluxetable*}{llll}
\tablecaption{Proposed heating mechanisms in plage or network regions}
\tablehead{
\colhead{Mechanism} & \colhead{Heating rate scaling} & \colhead{Location} & \colhead{Reference}
}
\startdata
%
Shock wave						&  $R^{{\rm a}} < 0$ with $B_{{\rm ch}}$\tablenotemark{b}		& inside flux tube			 		& e.g., \citet{cramer15} \\
Shock wave						&  												& near flux tube axis				 	& \citet{Hasan08} \\
Ohmic \& viscous heating 				& $R \sim 0$ with $ B_{{\rm ch}}$						& inside flux tube					& \citet{yadav21} \\
Shock wave						&  												& between flux tubes 	& \citet{snow18} \\
Alfv\'en wave turbulence 			& $R > 0$ with $ B_{{\rm corona}}$\tablenotemark{b}	 		& inside flux tube					& \citet{ballegooijen11} \\
Ohmic heating 						& $R > 0$ with curl $B_{{\rm ch}}$ 						& 								& e.g., \citet{rodriguez13a} \\
Ion - neutral collision					& $R > 0$ with $ B_{{\rm ch}}$							&					 				 		& \citet{soler17} \\
Ion - neutral collision					& $R > 0$ with $({\rm curl}\, B_{{\rm ch}})_{\perp}$\tablenotemark{c} & edges of flux tubes 						& \citet{kuridze16} \\
Ion - neutral collision					& $R > 0$ with $({\rm curl}\, B_{{\rm ch}})_{\perp}$			& inside flux tube	 				& \citet{shelyag16} \\
Ion - neutral collision					& $R > 0$ with $({\rm curl}\, B_{{\rm ch}})_{\perp}$			& edges of flux tubes	 	& \citet{khomenko18} \\
Ion - neutral collision					& $R > 0$ with $({\rm curl}\, B_{{\rm ch}})_{\perp}$			& fibrils			 	& \citet{khomenko18} \\
Magnetic reconnection 				& $R > 0$ with $B_{{\rm ch}}$							& above opposite $B_{{\rm ph}}$\tablenotemark{b} polarities & \citet{priest18} \\
\enddata
\tablenotetext{a}{$R$ is a correlation coefficient. $R > 0$ indicates a positive correlation.}
\tablenotetext{b}{$B_{{\rm ph}}$, $B_{{\rm ch}}$ and $B_{{\rm corona}}$ are magnetic field strengths in the photosphere, the chromosphere and the corona, respectively.}
\tablenotetext{c}{$\perp$ subscript denotes a component of curl $B_{{\rm ch}}$ perpendicular to the magnetic field direction.}
\label{table.mechanism1}
\end{deluxetable*}

Shock wave heating was first proposed by \citet{Schatzman49} as a mechanism of chromospheric heating.
In association with the characteristic pattern of shock waves found in temporal evolution of spectra, \citet{skogsrud16} observed small bright patches occurring in the transition region over a plage, and \citet{chintzoglou21} measured electron temperature enhancements in the chromosphere.
Having compared acoustic energy flux with radiative cooling energy flux using spectroscopic methods, \citet{sobotka16} and \citet{abbasvand2020a} claimed that the acoustic wave energy expected to be dissipated in the chromosphere by the shock waves is the dominant energy source to heat plage regions.
In contrast, \citet{abbasvand20b, abbasvand21} used the same technique and concluded that the acoustic energy deposited in plage regions is too low to balance the radiative loss, while the acoustic energy deposited in quiet Sun regions is a major contributor in balancing the radiative loss for the mid-chromosphere.
A definitive conclusion about shock wave heating, therefore, has not been reached observationally.

Shock waves may be generated by several mechanisms \citep{osterbrock61, hollweg71, suematsu82, kudoh99, pontieu04, shibata07, nakamura12, takasao13}.
The amplitude of upward propagating Alfv${\rm \acute{e}}$n waves excited by granule motions increases with the expansion of the flux tubes and becomes large enough to generate compressive shock waves that heat the chromosphere inside magnetic flux tubes \citep{matsumoto12, matsumoto14, wang20}.
Assuming that the magnetic field strength of the magnetic flux tubes is a constant, i.e., about 1 kG in the photosphere \citep{stenflo73, parker78, webb78, spruit79, nagata08, lagg10}, the strength of the shock waves is inversely proportional to the magnetic field strength in the chromosphere \citep[Table \ref{table.mechanism1}, ][]{cramer15}.
In a region consisting of multiple magnetic flux tubes, numerical simulations for weak shock waves repeatedly occurring with a short time scale less than 100 seconds yield chromospheric heating near the central part of the magnetic flux tube \citep[Table \ref{table.mechanism1}, ][]{Hasan08}.


Torsional motions have been observed in the chromosphere resulting from vortex flows in the photosphere (\citeauthor{brandt88} \citeyear{brandt88}; \citeauthor{wang95} \citeyear{wang95}; \citeauthor{bonnet08} \citeyear{bonnet08}; \\ \citeauthor{wedemeyer-bohm09} \citeyear{wedemeyer-bohm09}; \\ \citeauthor{wedemeyer-bohm12} \citeyear{wedemeyer-bohm12}).
Some numerical simulations show that torsional vortex motions can generate electric currents and shear flows inside flux tubes, which in turn heat the chromosphere via Ohmic heating and viscous dissipation \citep[Table \ref{table.mechanism1}, ][]{moll12, yadav21}.
In one of the simulation results, the horizontal structure of the chromospheric temperature does not correlate with the vertical component of the magnetic field in the chromosphere \citep[see Figure 6 of][]{yadav21}.
In addition, \citet{snow18} simulated two flux tubes independently perturbed by vortex motions in the photosphere. 
At the location where the flux tubes merge in the chromosphere, many small-scale transient magnetic substructures are generated from the interaction of the twisted or perturbed flux tubes.
Moreover, the interaction also amplifies the perturbations in the velocity and the magnetic field and produces upward propagating shocks that heat the chromosphere (Table \ref{table.mechanism1}).

Heating due to Alfv${\rm \acute{e}}$n wave turbulence is another candidate mechanism suggested for chromospheric heating \\ \citep{matsumoto12, liu14, ragot19}.
\citet{ballegooijen11} presented in their numerical test that the heating energy rate in the chromosphere increases with increasing magnetic field strength in the corona.
A steeper Alfv${\rm \acute{e}}$n wave velocity gradient enhances the reflection of Alfv${\rm \acute{e}}$n waves, creating stronger counter-propagating waves and correspondingly stronger turbulence in the chromosphere (Table \ref{table.mechanism1}).

Dissipation of magnetic energy by steady electric currents has also been suggested as a chromospheric heating mechanism \citep{parker83, rabin84, solanki03, hector05, tritschler08, morosin20}.
Magnetic flux concentrations expanding with height are associated with strong gradients in the magnetic field as well as a large current density \citep[Table \ref{table.mechanism1}, e.g., ][]{rodriguez13a, morosin20}.

Collisions between ions and neutral atoms enhance the dissipation of magnetic energy through various processes \citep{osterbrock61, goodman97, liperovsky00, krasnoselskikh10, khomenko12, sykora20}.
In general, the heating efficiency increases with the diffusion coefficient of the magnetic field due to collisions between ions and neutral atoms and the current density perpendicular to the magnetic field \citep[e.g.,][]{goodman04, khomenko18}.
As a first example, torsional Alfv${\rm \acute{e}}$n waves can be damped by collisions between ions and neutral atoms \citep[e.g., ][]{pontieu01, tu13, soler15b}.
\citet{soler17} performed numerical simulations and found that the chromospheric heating due to torsional Alfv${\rm \acute{e}}$n waves damped by collisions decreases with the increase of the expansion rate of the magnetic flux tube.
Assuming the magnetic field strength at the roots of the magnetic flux tubes is a constant in the photosphere \citep[e.g.,][]{stenflo73}, the expansion rate should be inversely proportional to the magnetic field in the chromosphere (Table \ref{table.mechanism1}).
Second, torsional or transverse Alfv${\rm \acute{e}}$n waves can also be damped by resonant absorption or by the Kelvin-Helmholtz instability \citep{okamoto15, antolin15, zaqarashvili15, giagkiozis16, terradas18}.
In chromospheric jets, collisions between ions and neutral atoms can play a dominant role in the diffusion of vortices produced at the boundary of the flux tubes by the Kelvin-Helmholtz instability and also in the dissipation of the magnetic energy \citep[Table \ref{table.mechanism1}, ][]{kuridze16}.
Third, collisions dissipate the magnetic energy, not only in Alfv${\rm \acute{e}}$n waves, but also for fast and slow magnetohydrodynamic waves and their multi-fluid counterparts \citep{soler13, soler15, shelyag16}. %
In this case, these additions open a multitude of pathways for heating dependent on the magnetic geometry.
 \citet{shelyag16} demonstrated that magnetic diffusion due to collisions between ions and neutral atoms leads to large and continuous dissipation of magnetic waves along the radius of a magnetic flux tube (Table \ref{table.mechanism1}).
In realistic three-dimensional radiative-magneto-hydrodynamic simulations, strong heating frequently occurs along inclined canopies in the photosphere or on fibril-like structures connecting opposite magnetic polarities \citep[Table \ref{table.mechanism1}, ][]{khomenko18}.


Finally, magnetic reconnection associated with magnetic flux cancellation or magnetic reconnection associated with horizontal magnetic fields has been suggested as sources of chromospheric heating \citep[e.g., ][]{isobe08, ishikawa09, priest18}.
In an analytical solution derived by \citet{priest18}, the heating energy rate is proportional to the magnetic field strength overlying the magnetic reconnection region (Table \ref{table.mechanism1}).


As Table \ref{table.mechanism1} summarizes, important discriminators of the possible heating mechanisms include the location of the heating and the correlation between the magnetic field properties in the chromosphere and the local heating rate.

A few observations already address the heating discriminants described in Table \ref{table.mechanism1}.
\citet{rodriguez13a} concluded that, in a plage region, the chromospheric temperature at the edges of magnetic flux tubes is similar to that near the flux tube axis from comparing observed intensity profiles of a chromospheric line, \ion{Ca}{2} 854 nm, with synthetic line profile computed from a numerical simulation \citep{gudiksen11}.
However, they did not derive which part is heated more strongly quantitatively.


\citet{pietrow20} and \citet{morosin20} measured vertically oriented fields with a field strength $\sim$ 400 G in chromospheric plage.
\citet{ishikawa21} inferred the line-of-sight component of the magnetic field strength at the top of the chromosphere in a plage region from a \ion{Mg}{2} spectro-polarimetric observation, and they presented a strong positive correlation between the field strength and the \ion{Mg}{2} line core intensity, which is a signature of chromospheric heating.
However, due to the low spatial resolution of these observations, it remains to be determined where the heating occurs on spatial scales comparable to individual magnetic flux tubes with a diameter of a few arcseconds in the chromosphere (e.g., see Figure 16 of \cite{morosin20}).
The positive correlation might also result from the global structure of the plage region, rather than individual small heating events.

Our aim is to investigate the relationship between the signatures of heating in a plage region and the magnetic field in the photosphere and the chromosphere using spectro-polarimetric observations 
to identify what mechanisms are at work.
In order to determine the heating mechanisms, we must be able to infer the spatial distribution of the magnetic field strength and electric current density on a sub-arcsecond spatial scale ($\sim 500 \, {\rm km}$), which is fine enough to clarify where heating occurs locally and in relation to individual magnetic flux tubes \citep{morosin20} in the plage chromosphere.
We also have strong temporal constraints due to rapid dynamic events that affect the formation of chromospheric lines; e.g., spicules evolve on a temporal sclae of ${\sim}$10 seconds \citep{hansteen06, pontieu07, anan10}.
Therefore, we need to measure Stokes spectra in two spatial dimensions covering magnetic flux tubes within ${\sim}$10 seconds.
Integral field spectro-polarimeters are the most suitable instruments to meet these requirements \citep[e.g., ][]{kleint17}.


\section{Observations} \label{sec:observation}

\begin{figure*}
\begin{center}
\includegraphics[angle=0,scale=1.0,width=160mm]{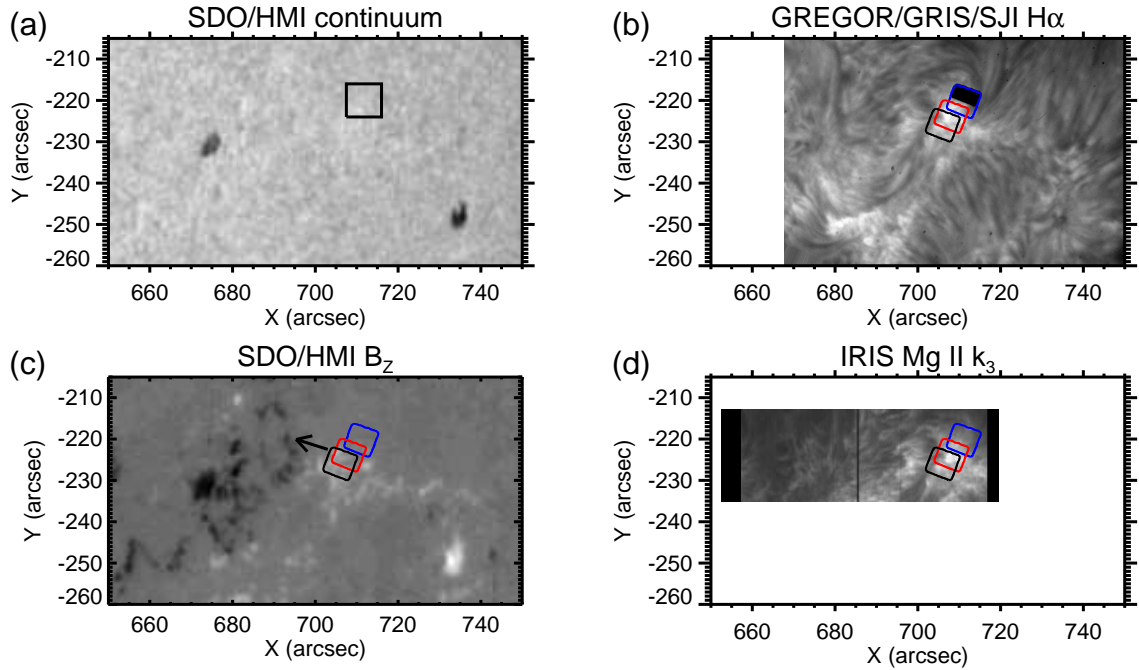}
\end{center}
\caption{
Active region NOAA 12723 observed at 9:38 UT in (a) 617 nm continuum intensity taken by SDO/HMI, (b) H${\rm \alpha}$ intensity taken by a slit jaw imager of the GREGOR/GRIS, (c) the vertical component of the magnetic field in the photosphere measured by SDO/HMI, and (d) the \ion{Mg}{2} k line core intensity taken by the IRIS spectrograph.  
The black square in panel (a) marks a region enlarged in Figure \ref{fig.obs_enlarge}.
Black, red, and blue oblique boxes in panels (b) - (d) indicate three regions (Plage A, B, and C in this paper) that GRIS observed.
The black rectangular shadow inside the blue oblique box in panel (b) corresponds to an aperture of the field stop for a single position of the GRIS IFU.
The black arrow in panel (c) points to the direction of to the solar disk center.
		}
\label{fig.obs_ar}
\end{figure*}

A plage region in NOAA active region 12723 was observed between 9:25 and 9:58 UT on 2018 October 3 with the $1.5$ m GREGOR solar telescope \citep{vonderluhe01, schmidt12} on Tenerife in the Canary Islands, cooperatively with the Interface Region Imaging Spectrograph \citep[IRIS, ][]{pontieu14} spacecraft and the Helioseismic and Magnetic Imager \citep[HMI, ][]{scherrer12} onboard the Solar Dynamics Observatory \citep[SDO, ][]{pesnell12} spacecraft.
The use of the adaptive optics system mounted on GREGOR \citep{berkefeld12, berkefeld18} improves image quality, which is blurred and distorted by the Earth's atmosphere. 
During the observation, we measured a Fried's seeing parameter \citep{fried66} of $6.2 \, \pm \, 2.1$ cm.

Figure \ref{fig.obs_ar} shows images of the active region taken at 9:38 UT.
The region was bipolar with two pores which had opposite polarities (panels a and c).
Around the pores, there were plage regions, which are bright in chromospheric lines, e.g., the \ion{H}{1} 656 nm (H${\rm \alpha}$) and the \ion{Mg}{2} k lines (panels b and d).

Stokes spectra in a near infrared spectral window containing the \ion{He}{1} triplet and \ion{Si}{1} 1082.7 nm were obtained by the GREGOR Infrared Spectrograph \citep[GRIS, ][]{collados12}, which uses an image-slicer type Integral Field Unit (IFU) \citep[][; Dominguez-Tagle et al. in prep]{tagle18}. 
The IFU enables measurement of Stokes spectra in two spatial dimensions with a field of view of $6.075 \times 3.0\, {\rm arcsec^2}$.
The rectangular field of view is sliced by 8 slitlets each having dimensions $6.075 \times 0.375\, {\rm arcsec^2}$.
In the raw data, the spectral images from each slitlet are clearly separated by dark bands.
The first step of the reduction process identifies their position and removes the dark bands from all files (map, dark, flatfield and polarimetric calibration files), creating a continuous spectral image in the spatial direction as if it were coming from a continuous long slit.
From there on, the standard procedure for dark current and flat field correction was applied as described in \citet{schlichenmaier02} for the Tenerife Infrared Polarimeter \citep[TIP, ][]{collados99, pillet99} data.
Calibration optics installed at the secondary focus, before any oblique reflection, were used to measure the demodulation matrix at the time of the calibration.
A model of the telescope was used to update the modulation matrix to the time of the observations, updating it at every individual scan position. The polarization model of the telescope uses as free parameters the refractive indices of the mirrors, which are obtained after calibration data taken with the telescope pointing at around 10 different orientations on the sky over the span of multiple days. As the last reduction step, the spectral images of the dual-beam setup are combined, and a cross-talk correction from ${\it I}$ to ${\it Q}$, ${\it U}$ and ${\it V}$ is calculated and applied to force the continuum polarization to zero for each individual Stokes spectrum.
Finally, the calibrated spectral images corresponding to each slice are separated and put in their corresponding position and time in the measured 2 dimensional field of view, creating four-dimensional data cubes of ${\it I}$, ${\it Q}$, ${\it U}$, ${\it V} (\lambda, x, y, t)$.

During this observation, the solar image was scanned across the IFU using 2 steps in the direction parallel to the short side of the IFU resulting in a total field of view of $6.075 \times 6.0 \, {\rm arcsec^2}$.
The two-dimensional spatial distributions of the full Stokes spectra were sequentially obtained with a temporal cadence of 26 seconds.
In total we observed three regions marked by oblique boxes in Figure \ref{fig.obs_ar} panels (b) - (d): 11 maps of Plage A (black box), 13 maps of Plage B (red box) and 48 maps of Plage C (blue box).
To increase the sensitivity, we binned the data in both the spatial (i.e. along the slit direction) and spectral dimensions to 0.276 arcsec pixel$^{-1}$ and 5.4 pm pixel$^{-1}$, respectively.  The final polarization sensitivity, defined here as the standard deviation of the locally normalized continuum intensity, are $\sim 0.6 \%$, $\sim 0.03 \%$, $\sim 0.04 \%$, and $\sim 0.03 \%$, for {\it I}, {\it Q}, {\it U}, and {\it V},  respectively.

\begin{figure}
\begin{center}
\includegraphics[angle=0,scale=1.0,width=80mm]{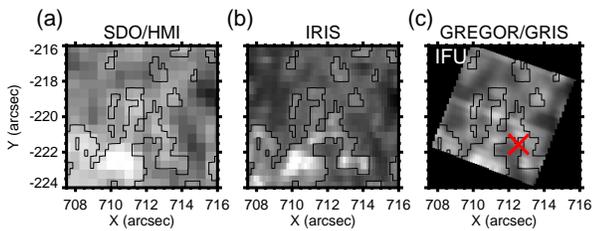}
\end{center}
\caption{
	Continuum intensity maps taken by (a) the SDO/HMI, (b) the IRIS spectrograph, and (c) the GREGOR/GRIS IFU mode. All contour lines are drawn with reference to the map in panel (b). The red cross marks where profiles in Figure \ref{fig.stokes_profiles} were observed. The GRIS field of view is shown by the blue box in panel (b) of Figure \ref{fig.obs_ar}, i.e., Plage C. 
		}
\label{fig.obs_enlarge}
\end{figure}

Figure \ref{fig.obs_enlarge} shows enlarged images of the plage region inside the black square in panel (a) of Figure \ref{fig.obs_ar}, which includes the GRIS field of view for Plage C.
GRIS and IRIS maps were co-aligned with HMI observations by comparing features in the continuum images (see the black contour lines in Figure \ref{fig.obs_enlarge}).
A two-channel slit-jaw imager provides context imaging for the GRIS IFU with filters to cover the 770 nm continuum and H${\rm \alpha}$ line core.
A reflective field stop placed at the entrance of the IFU directs light to the slit-jaw system, while the light passing through a rectangular hole in the field stop reaches the image slicer.
Figure \ref{fig.obs_ar}(b) shows an image of the active region taken by the H${\rm \alpha}$ imager.

While GRIS observed the plage region, the IRIS spacecraft scanned the active region including the plage region nine times (OBSID: 3600504455).
The scan step was 0.35 arcsec, and the raster scan cadence was 204 seconds.
Spectra of the ultraviolet \ion{Mg}{2} h \& k doublet as well as the ultraviolet continuum near 283 nm were obtained by IRIS at the same time.
They were calibrated with dark subtraction, flat-field correction, and geometric and wavelength calibration.
The \ion{Mg}{2} k intensity map and the ultraviolet continuum intensity map are shown in panel (d) of Figure \ref{fig.obs_ar} and in panel (b) of Figure \ref{fig.obs_enlarge}, respectively. 

\section{Analysis} \label{sec:method}

We derived the magnetic field in the photosphere and the chromosphere in addition to the radiative cooling energy flux of the \ion{Mg}{2} h \& k emissions in order to investigate the association between the magnetic field structure and the chromospheric heating. 
In the chromosphere, the heating can be assumed to be balanced by the radiative cooling.

We analyzed maps taken when both GRIS and IRIS scanned the same region at the same time.
The number of the maps that satisfy this condition for Plage A, B, and C are 1, 2, and 6, respectively.
As the H${\rm \alpha}$ and the \ion{Mg}{2} k images show (Figure \ref{fig.obs_ar}), Plage A and B fully overlap the bright plage region, while Plage C was at the edge (or periphery) of the plage region.

\subsection{Magnetic fields}
\label{sec:method:b}

\subsubsection{Inversion}
\label{sec:method:b:inversion}

\begin{figure}
\begin{center}
\includegraphics[angle=0,scale=1.0,width=80mm]{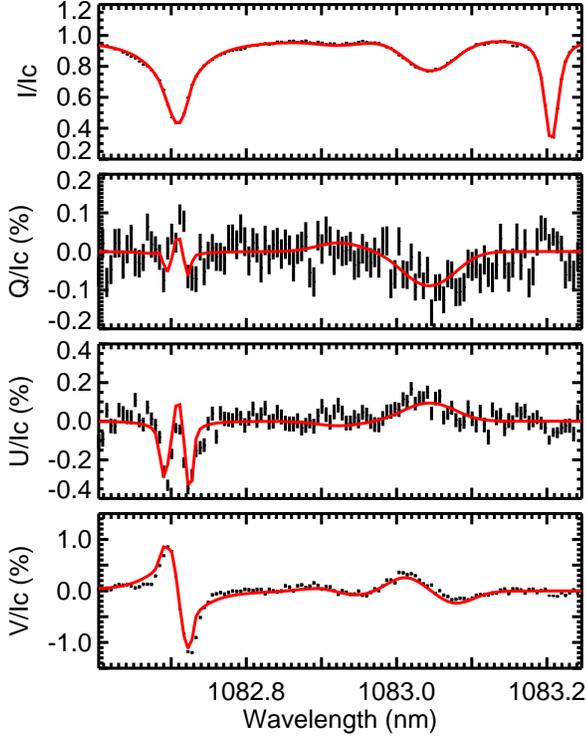}
\end{center}
\caption{
Observed Stokes profiles at the location marked by the red cross in panel (c) of Figure \ref{fig.obs_enlarge} points (black) and an inversion result (red).
		}
\label{fig.stokes_profiles}
\end{figure}

An example full Stokes spectrum obtained by GRIS is shown in Figure \ref{fig.stokes_profiles}.
In order to fit all the observed spectra and infer physical quantities, we applied the HAnle and ZEeman Light v2.0 inversion code \citep[HAZEL2, ][]{asensio08} to the Stokes spectra of the \ion{Si}{1} 1082.7 nm and the \ion{He}{1} 1083.0 nm (the red lines in Figure \ref{fig.stokes_profiles}).
HAZEL2 carried out three optimization cycles using weights summarized in Table \ref{table.weight}.
At the same time, a telluric line of ${\rm H_{2}O}$ at 1083.211 nm was fitted using a Voigt function. 

\begin{table}
\begin{center}
\caption{Weights for all Stokes parameters}
\begin{tabular}{crrrrrrrrrrr}
\tableline\tableline
          & \multicolumn{4}{c}{ \ion{Si}{1}} & \multicolumn{4}{c}{ \ion{He}{1}} \\
Cycle & {\it I} & {\it Q} & {\it U} & {\it V} & {\it I} & {\it Q} & {\it U} & {\it V} \\ 
\tableline
1 & 3 & 0 & 0  & 0 &  2 & 0  & 0 & 0 \\
2 &  0 & 1 & 1  & 1  & 0 & 1   & 1  & 1 \\
3 & 3 & 2 & 2  & 2 & 2 & 2 & 2 & 2 \\
\tableline
\end{tabular}
\end{center}
\label{table.weight}
\end{table}

Inversions of the \ion{Si}{1} Stokes spectra, which are formed in the upper photosphere \citep{bard08, shchukina17}, were performed by HAZEL2 using Stokes Inversion based on Response functions \citep[SIR, ][]{cobo92} in local thermodynamic equilibrium.
The atmospheric model consists of physical variables specified at a number of fixed-height nodes.  Interpolation between nodes provides the variation of each quantity with height, and perturbing the value at each node leads to the best-fit model.
We used 5 nodes for temperature, 2 nodes for bulk velocity, and assumed that both the magnetic field and micro-turbulent velocity are uniform with height.
With this assumption, the inferred magnetic field can be interpreted as an average over the line of sight in the photosphere.
We also assume the filling factor is equal to 1.

\begin{figure}
\begin{center}
\includegraphics[angle=0,scale=1.0,width=80mm]{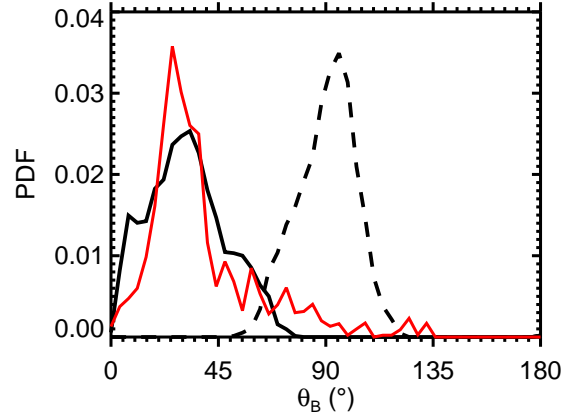}
\end{center}
\caption{
Probability density functions (PDF) of the inclination angle of the magnetic field in the photosphere with respect to the local solar vertical.
The black solid and dashed lines show PDFs of two candidates for the solution of the inversion applied to the GRIS data, and the red line displays results from SHARPs.
}
\label{fig.thb_histogram}
\end{figure}

All magnetic fields inferred from the inversion for the \ion{Si}{1} Stokes profiles, whose signals are dominated by the Zeeman effect, suffer from a $180^{\circ}$ azimuthal ambiguity in a line-of-sight reference frame, $\Phi_{B}$, \citep[e.g.,][]{landi04}.
We derived the other candidate solution from the one inferred from the inversion by changing $\Phi_{B}$ by $180^{\circ}$.
We then classified the two candidates into the vertical and the horizontal ones in the reference frame of the local solar vertical.
In Figure \ref{fig.thb_histogram}, we compare the histograms of the two classes of solutions, across the field-of-view and in the local vertical reference frame, with that found in the associated Space-weather HMI Active Region Patches (SHARPs) dataset \citep{bobra14}.
The azimuthal ambiguity has been resolved in the SHARP data using a minimum energy method \citep{metcalf94, leka09}.
We select the vertical class of solutions as the true solution based on their favorable overall agreement with the SHARP data.
The finding that most of the magnetic field vector in the photosphere was almost perpendicular to the solar surface over the plage region is consistent with that reported by \cite{pillet97} and \cite{buehler15}.

The \ion{He}{1} 1083.0 nm Stokes profiles were fitted taking into account all relevant physical mechanisms producing or modifying polarization signals in the spectral line: the Zeeman effect, the Paschen-Back effect, scattering polarization, and the Hanle effect.
The atmospheric model consists of a constant property slab located above the solar surface at a height of 2 arcsec.
The slab is characterized by the magnetic field vector, an optical depth of the line, the Doppler broadening, the damping parameter of the Voigt absorption profile, and the bulk velocity.   

Magnetic fields inferred from the \ion{He}{1} triplet inversion in the Hanle saturated regime are subject to $90^{\circ}$ and $180^{\circ}$ azimuth ambiguities described by \citet{merenda06}, \citet{asensio08}, and \citet{schad13}.
Which ambiguities influence an inversion result depends on geometrical configurations of the actual magnetic field orientation and the location of the observed region on the solar disk.
For example, Figure 3 of \citet{schad15} shows regions of inversion ambiguities over an active region. In the south of the sunspot that they observed, the inferred magnetic fields are affected by all $+90^{\circ}$, $-90^{\circ}$ and $+180^{\circ}$ ambiguities, meanwhile some regions are unaffected by ambiguities.
We generated four solution candidates for each inversion and we use the following algorithm to narrow down the solution space: 
\begin{enumerate}
\item run the first inversion described above,
\item run three more inversions using fixed thermal parameters determined in the inversion at point 1 as well as fixing $\Phi_{B} = \{ \Phi_{B, 1} + 90^{\circ}$, $\Phi_{B, 1} - 90^{\circ}$, $\Phi_{B, 1} + 180^{\circ} \}$, where $\Phi_{B, 1}$ is given by the first inversion,
\item for each of the four inversions, calculate the sum of squared residuals in Stokes ${\it Q}$ and ${\it U}$ in a wavelength range between the spectral line center $\pm$ the inferred line width,
\item identify the best fit solution that has the smallest sum of squared residuals, and
\item select candidates where the difference of the sum of squared residuals from the smallest one are less than the square of the observational noise, in addition to the best fit solution.
\end{enumerate}

\begin{figure}
\begin{center}
\includegraphics[angle=0,scale=1.0,width=80mm]{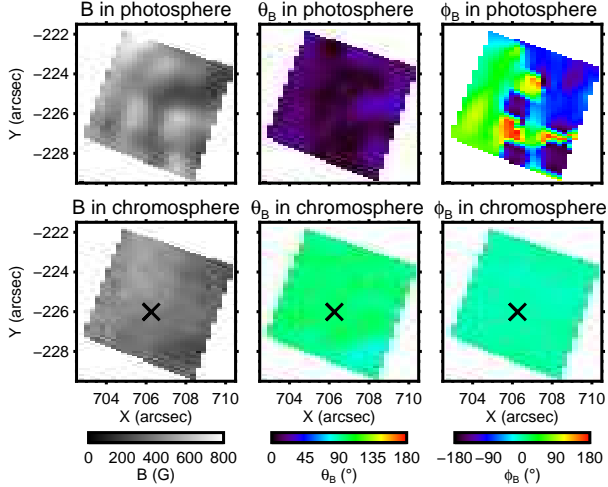}	
\end{center}
\caption{
Spatial distributions in Plage A of the strength (left), the inclination angle (middle), and the azimuth angle (right) of the inferred magnetic field vector at the photosphere (top row) and the chromosphere (bottom row) in the reference frame of the local solar vertical.
The reference direction for the azimuth angles is towards the solar disk center.  
The black crosses mark the pixel where Stokes profiles used for Figure \ref{fig.chi} were observed.
		}
\label{fig.b_map}
\end{figure}

\begin{figure}
\begin{center}
\includegraphics[angle=0,scale=1.0,width=80mm]{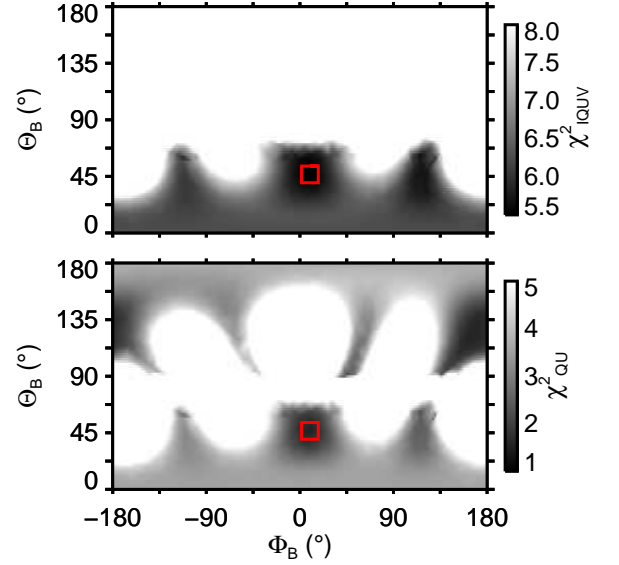}
\end{center}
\caption{
(Top) Mean square of residuals normalized by noise level in full Stokes parameters, $\chi^2_{{\it IQUV}}$, and (bottom) in Stokes ${\it Q}$ and ${\it U}$, $\chi^2_{{\it QU}}$, as functions of $\Theta_{B}$ and $\Phi_{B}$ in the chromosphere.
The red squares denote the solution of the inversion.
		}
\label{fig.chi}
\end{figure}

Figure \ref{fig.b_map} shows the inferred magnetic field strength, $B$, its inclination angle with respect to the local solar vertical, $\theta_{B}$, and its azimuth angle with respect to the direction to the solar disk center, $\phi_{B}$.
For the chromosphere, most of the Stokes spectral profiles have a unique solution without ambiguities.
It might correspond to the case reported by \citet{schad15}, in which there are some regions where the inferred magnetic field does not have any ambiguity.
To confirm the uniqueness of the solution, we performed 2000 inversions for a Stokes profile observed at the location marked by the cross in Figure \ref{fig.b_map} with chromospheric angles for $\Theta_{B}$ and $\Phi_{B}$ fixed to randomly distributed values in the ranges $0^{\circ} < \Theta_{B} < 180^{\circ}$ and $-180^{\circ} < \Phi_{B} < 180^{\circ}$.
Figure \ref{fig.chi} shows the mean square of residuals normalized by the noise level. 
It demonstrates that the observed Stokes ${\it V}$ excludes a magnetic field vector that has $\Theta_{B} > 90^{\circ}$ from the solution.
Moreover, for Stokes ${\it Q}$ and ${\it U}$, the residuals of the best-fit solution are significantly smaller than those in the other local minima.

The inclination angle and the azimuth angle maps for the chromosphere display a relatively uniformly distributed horizontal magnetic field primarily directed towards the solar disk center (the direction indicated by the black arrow in panel c of Figure \ref{fig.obs_ar}), while for the photosphere the magnetic field was almost vertical to the solar surface (Figure \ref{fig.thb_histogram} and \ref{fig.b_map}).
 The measured change in direction of the magnetic field between the photosphere and chromosphere is broadly consistent with our expectations for this region, where the field should rapidly turn over with height to connect to the nearby opposite polarity across the large scale diffuse polarity inversion line around the head of the black arrow in Figure \ref{fig.obs_ar}c.
The dark threads connecting across the polarity inversion line observed in both H${\rm \alpha}$ and the \ion{Mg}{2} k support this interpretation (panels b and d of Figure \ref{fig.obs_ar}).
Note that for a comparable set of observations both \citet{pietrow20} and \citet{morosin20} found that the field remained mostly vertical, but their observations were situated very close to pores.

Finally, we calculated the electric current density $ \vec{J} = (4 \pi / c) \, \vec{\nabla} \times \vec{B}$ using the inferred magnetic field vectors in the photosphere and chromosphere.
For the derivative in the $X$-$Y$ plane, which is horizontal to the solar surface, a central difference scheme with the truncation error of $O(\Delta x^2)$ was used.
For the calculation of $\partial B_{Z} / \partial X$ or $\partial B_{Z} / \partial Y$, we took the average of $B_{Z}$ in the photosphere and $B_{Z}$ in the chromosphere taking into account the apparent displacement derived in section \ref{sec:method:align} before evaluating the derivative. 
For the derivative in the vertical direction, a height difference between the photosphere and the chromosphere of 2 arcsec ($=1450$ km) was assumed by considering the apparent displacement.

\subsubsection{Inversion errors}
\label{sec:method:b:err}

We investigated inversion errors by applying HAZEL2 to 10,000 synthetic Stokes profiles, which were calculated with randomly distributed parameters of the magnetic field and the line-of-sight velocity, $V_{{\rm LOS}}$, and to which random noise with variances matching the observational noise were added.
For the \ion{Si}{1} inversions, the considered range of $B$, $\Theta_{B}$, $\Phi_{B}$, and $V_{{\rm LOS}}$ are, respectively, $0\,{\rm G} < B < 1000\,{\rm G}$, $0^{\circ} < \Theta_{B} < 180^{\circ}$, $-180^{\circ} < \Phi_{B} < 180^{\circ}$, and $-10\,{\rm km \, s^{-1}} < V_{{\rm LOS}} < 10\,{\rm km \, s^{-1}}$, where $\Theta_{B}$ is the inclination angle of the magnetic field with respect to the line of sight.

\begin{figure}
\begin{center}
\includegraphics[angle=0,scale=1.0,width=80mm]{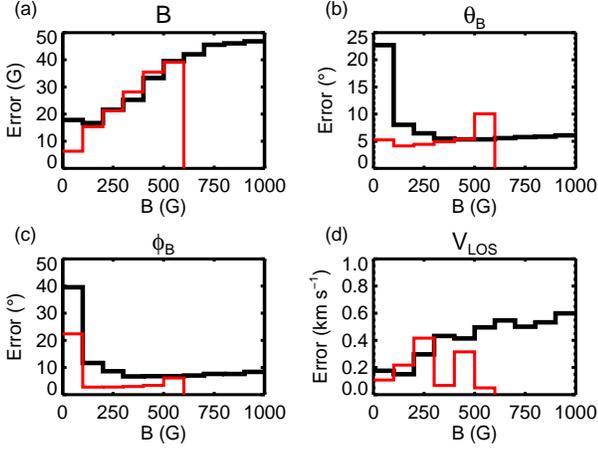}
\end{center}
\caption{
	Root mean squares of errors as a function of the magnetic field strength in (a) the magnetic field strength, (b) the inclination and (c) the azimuth angles of the magnetic field with respect to the solar vertical, and (d) the line-of-sight component of the velocity. Black and red lines indicate the errors for the photosphere and the chromosphere, respectively.
		}
\label{fig.cal_error}
\end{figure}

Figure \ref{fig.cal_error} shows the root mean square of the difference between the inferred values and the values used for synthesizing Stokes profiles. 
The error in the magnetic field strength increases with the field strength.
For the range of field strengths inferred in the observed plage region ($\sim 100$ to 800 G in Figure \ref{fig.b_map}), the error in the photospheric inclination and azimuth angles are approximately equal to 5$^{\circ}$.
The error in the line-of-sight velocity is less than 0.5 km s$^{-1}$.

\begin{figure}
\begin{center}
\includegraphics[angle=0,scale=1.0,width=80mm]{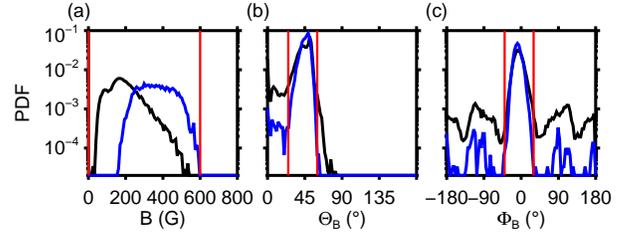}
\end{center}
\caption{
	PDF of (a) the strength, (b) the inclination angle and (c) the azimuth angle of the magnetic field in the chromosphere in the line-of-sight reference frame for plage region (blue) and its periphery (black). The red vertical lines indicate the range used in the numerical test for the inversion error.
		}
\label{fig.pdf_chb}
\end{figure}

For the \ion{He}{1} inversions, since the sensitivity of the linear polarization to changes of the magnetic field strongly depends on the magnetic field, we narrowed down the range for the parameters of the synthetic Stokes spectra. 
Figure \ref{fig.pdf_chb} shows probability density functions for chromospheric $B$, $\Theta_{B}$ and $\Phi_{B}$ inferred from the GRIS data.
The narrowed ranges were determined to be $0\,{\rm G} < B < 600\,{\rm G}$, $25^{\circ} < \Theta_{B} < 60^{\circ}$ and $-40^{\circ} < \Phi_{B} < 30^{\circ}$, such that $98\%$ of the parameters vary within these ranges for the plage region and $83\%$ for its periphery.
The range of $V_{{\rm LOS}}$ are the same as those for the \ion{Si}{1} inversions above.
The errors in the magnetic field strength, the inclination, the azimuth angles, and the line-of-sight velocity are almost same as those for the \ion{Si}{1} inversions (Figure \ref{fig.cal_error}).

\begin{figure}
\begin{center}
\includegraphics[angle=0,scale=1.0,width=80mm]{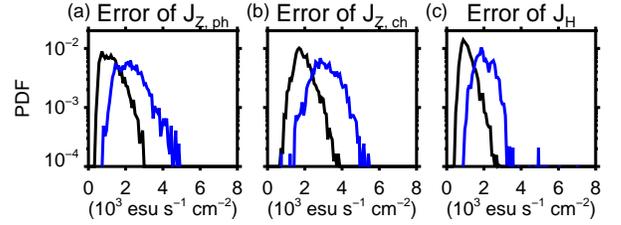}
\end{center}
\caption{
	PDF of (a) the $J_{{\rm Z, ph}}$ error, (b) the $J_{{\rm Z, ch}}$ error, and (c) the $J_{{\rm H}}$ error for plage region (blue) and its periphery (black).
	}
\label{fig.pdf_error_current}
\end{figure}

In order to evaluate errors of the electric current density, we calculated the current density $100$ times after first adding random noise to the inferred magnetic field vectors with a standard deviation equal to the errors derived above.
Because the formation height of the He line can spatially change with a peak-to-valley of $2$ Mm \citep{leenaarts16}, we also add this variance in the height difference between the photosphere and chromosphere as a source of error. 
We set the variance to be $0.3$ Mm in the calculations.
Figure \ref{fig.pdf_error_current} shows histograms of the standard deviation of the $100$ calculation results for each pixel for vertical components of the current density in the photosphere, $J_{{\rm Z, ph}}$, and in the chromosphere, $J_{{\rm Z, ch}}$, and its horizontal component, $J_{{\rm H}} = \sqrt{J_{X}^{2} + J_{Y}^{2}}$.
We found typical current densities of $5 \times 10^3$ esu s$^{-1}$ cm$^{-2}$, while typical errors of $J_{{\rm Z, ph}}$, $J_{{\rm Z, ch}}$ and $J_{{\rm H}}$ for the plage region are $2 \times 10^3$, $3 \times 10^3$, and $2 \times 10^3$ esu s$^{-1}$ cm$^{-2}$, respectively, and for the periphery region $1 \times 10^3$, $2 \times 10^3$, and $1 \times 10^3$ esu s$^{-1}$ cm$^{-2}$, respectively.

\subsection{\ion{Mg}{2} h \& k flux}
\label{sec:method:mg2}

\begin{figure}
\begin{center}
\includegraphics[angle=0,scale=1.0,width=80mm]{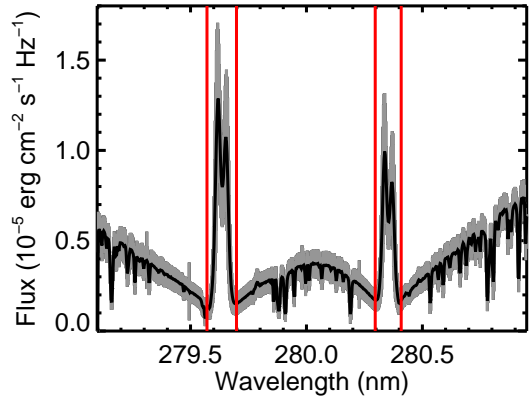}
\end{center}
\caption{
\ion{Mg}{2} h \& k flux profile spatially averaged over the regions of interest.
Gray bars display the standard deviation in the flux at each wavelength.
Red lines indicate spectral ranges used to estimate total integrated radiative cooling energy flux of the spectral lines.
		}
\label{fig.mg_profile}
\end{figure}

The ultraviolet \ion{Mg}{2} doublet typically appears as two broad absorption lines with self-reversed emission cores in solar spectra emitted from outside of sunspots \citep[e.g. ][]{durand49, lemaire73}.
Figure \ref{fig.mg_profile} shows a spatially averaged \ion{Mg}{2} h \& k flux profile over the regions of interest.
The two self-reversed emission cores of the \ion{Mg}{2} k and h appear at wavelengths of 279.6 nm and 280.3 nm, respectively.
The flux was estimated using the $iris\_calib.pro$ routine, part of the IRIS SolarSoftware tree \citep{freeland98}.
The routine derives the flux using the observed counts (in the data number units) with some information at the time of the observation, e.g., the number of photons per data number, the effective area, the size of the spatial pixels, the size of the spectral pixel, the exposure time, and the slit width.

We obtained the total radiative cooling energy flux of the \ion{Mg}{2} h \& k emissions by integrating the flux with respect to the wavelength in two ranges between 279.571 nm and 279.698 nm and between 280.296 nm and 279.698 nm (the red vertical lines in Figure \ref{fig.mg_profile}).
The error of the total radiative cooling energy flux was estimated to be about $0.6$ \% from the photon noise and the camera readout noise \citep{pontieu14}. 


\subsection{Alignment of chromospheric maps with photospheric ones}
\label{sec:method:align}

We investigate the structure of the magnetic field not only in the chromosphere but also in the underlying photosphere.
However, because the plage region was observed close to the solar limb ($\mu \sim 0.65$), the projection effect is significant.
In addition, the horizontal magnetic field in the chromosphere implies that the magnetic flux tubes were bent or inclined toward the disk center.
Therefore, we have to consider apparent spatial displacements in order to investigate the spatial correspondence between photospheric structures and strong chromospheric heating.
For the spatial correspondence between structures and heating in the chromosphere, we assume that the formation height of the \ion{He}{1} line is the same as those of the \ion{Mg}{2} h \& k lines.

To derive the apparent displacements, we used spatial distributions of \ion{Mg}{2} k intensity formed at different heights in the atmosphere.
The formation of the \ion{Mg}{2} h \& k lines has been studied theoretically (\citeauthor{athay68} \citeyear{athay68}; \citeauthor{milkey74} \citeyear{milkey74}; \citeauthor{gouttebroze77} \citeyear{gouttebroze77}; \\ \citeauthor{lemaire83} \citeyear{lemaire83}; \citeauthor{uitenbroek97} \citeyear{uitenbroek97}; \\ \citeauthor{leenaarts13a} \citeyear{leenaarts13a}).
The intensity in a central depression of the self-reversed profile, which is often referred to as \ion{Mg}{2} h3 or k3, is formed in the upper chromosphere or transition region \citep[e.g., ][]{vernazza81}. 
In addition, it corresponds to a geometrical height at which the optical depth is equal to unity at the line core wavelength \citep{leenaarts13b}.
The intensity of the two emission peaks in the line core, which are often referred to as \ion{Mg}{2} h2 or k2, corresponds to the temperature in the middle chromosphere \citep{leenaarts13b}.
Because departure from LTE is small in the atmosphere below the formation heights of Mg h2 and k2, the intensity profile from the peak to the line wing displays temperature variation along the line of sight from the middle chromosphere to the photosphere.
The side dips of the self-reversed profiles, often referred to as \ion{Mg}{2} h1 or k1, are formed near the temperature minimum \citep[e.g., ][]{ayres76}.

\begin{figure}
\begin{center}
\includegraphics[angle=0,scale=1.0,width=80mm]{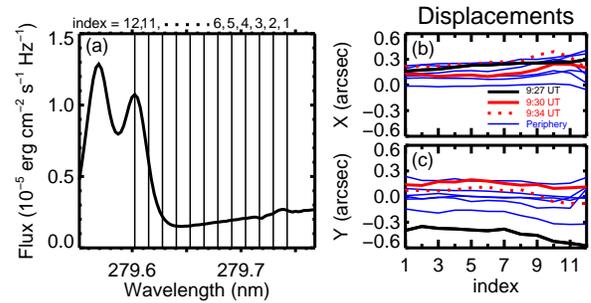}
\end{center}
\caption{
	(a) Indexes in wavelength indicated by vertical solid lines on the averaged \ion{Mg}{2} k flux profile, (b) accumulated displacements of intensity maps from the continuum image (283 nm) in $X$ direction, and (c) in $Y$ direction.
	The wavelength for each intensity map is shown in the horizontal axis as the index.
	Black, red, and blue lines show the displacements for Plage A, Plage B, and Plage C, respectively. IRIS had scanned Plage C 6 times with the cadence of 204 seconds since 9:38 UT.
		}
\label{fig.projection}
\end{figure}

\begin{figure}
\begin{center}
\includegraphics[angle=0,scale=1.0,width=80mm]{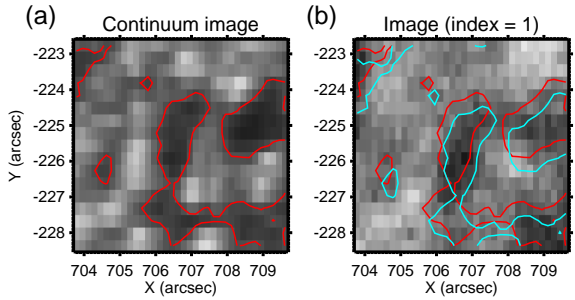}
\end{center}
\caption{
	(a) Continuum intensity map and (b) intensity map of \ion{Mg}{2} k line-wing, of which the wavelength is indicated by index 1 in Figure \ref{fig.projection}.
	The red contour lines in both maps are drawn with reference to the continuum intensity map.
	The blue contour is the red one shifted by an offset derived with a cross correlation technique.
		}
\label{fig.projection2}
\end{figure}

Because thermal and magnetic structures in the chromosphere are too different from those in the photosphere to 
compare directly, we calculated apparent spatial displacements between intensity maps in successive wavelength bands, moving from the \ion{Mg}{2} k line wing to line core, marked by vertical lines in Figure \ref{fig.projection}(a), and the continuum image at 283 nm, which is relatively free of spectral lines.
Each wavelength approximately corresponds to thermal structures at successively greater heights between the photosphere and chromosphere.
At first, a rigid displacement was derived using a cross correlation technique between the intensity maps of the continuum and \ion{Mg}{2} k line wing at the location indicated by index 1 in Figure \ref{fig.projection}.
Figure \ref{fig.projection2} shows that the spatial alignment of the line-wing-intensity map with the continuum intensity map shifted by the derived offset is better than that with the original continuum intensity map in the center portion of the field of view.
Next, we iteratively derived rigid displacements between intensity maps with indexes $i$ and $i+1$ as $i$ ranged from $1$ to $11$.
Figure \ref{fig.projection}(b) and (c) show the inferred apparent displacements of an intensity map from the continuum image in each wavelength.

The field of view of the analyzed intensity maps is $6 \times 6$ arcsec$^{2}$ and the center of the field is the same as that of the IFU observation.
In order to increase the signal-to-noise ratio, intensities were integrated with respect to wavelength with a spectral range of 0.013 nm, which corresponds to the interval of the vertical lines in Figure \ref{fig.projection}.
Variation in the lines of the same color, e.g., red or blue, reflect the sensitivity of this method, because the spatial samplings of IRIS in $X$ and $Y$ are 0.17 and 0.35 arcsec/pixel, respectively.

The total apparent displacements from the continuum images to \ion{Mg}{2} k2 tend to be positive in X ($0.0$ - $0.3$ arcsec, i.e., $0$ - $200$ km), which is consistent with the direction of the displacement caused by the projection effect.
On the other hand, there are no significant displacements in Y except for one indicated by the black line ($-0.6$ arcsec, i.e., $-450$ km).

\begin{figure*}
\begin{center}
\includegraphics[angle=0,scale=1.0,width=160mm]{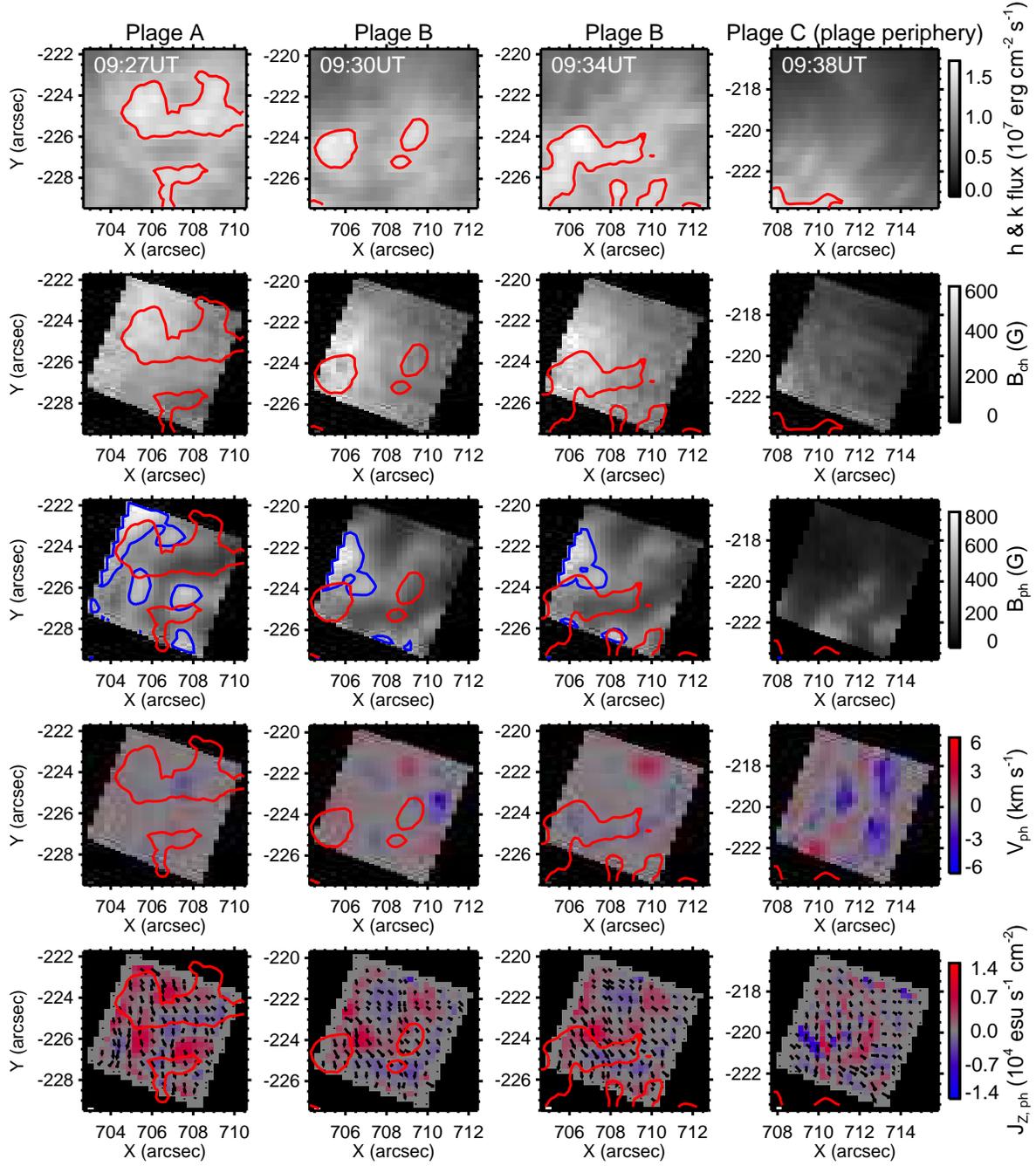}	
\end{center}
\caption{
Spatial distributions of the total integrated radiative cooling energy flux of the \ion{Mg}{2} h \& k lines, chromospheric magnetic field strength, photospheric magnetic field strength, photospheric line-of-sight velocity, and the solar vertical component of the electric current density in the photosphere (from top to bottom).
Different columns show those taken at different times.
Red and blue contour lines indicate where the \ion{Mg}{2} h \& k total radiative energy flux and $B_{{\rm ph}}$ are equal to $1.35 \times 10^{7} {\rm \,erg\,cm^{-2}\,s^{-1}}$ and $500$ G, respectively.
Black arrows in the bottom plot indicate the solar horizontal component of the electric current density.
		}
\label{fig.result1}
\end{figure*}
%
%
%

The coalignment strategy described above allows us to directly compare \ion{Mg}{2} h \& k emissions with the photospheric magnetic field using the cumulative displacement between maps of the continuum intensity and peak \ion{Mg}{2} intensity (see Figure \ref{fig.result1} rows 3-5 and subsequent analysis, below).
This procedure essentially assumes that all misalignment between thermal structures at different heights acts like projection effects, but in reality there is degeneracy between thermal structure, magnetic structure, and projection effects.
The observed magnetic field has both varying inclination, and changes in inclination with height, throughout the FOV.
In addition, thermal structures do not necessarily represent magnetic structures \citep{rodriguez11, schad13, leenaarts15, sykora16, asensio17}.
Therefore, interpreted in the most conservative fashion, we can derive general relationships between the magnetic field in the photosphere and chromospheric heating, but not individual relations.
At the same time, because the chromospheric field has a relatively uniform orientation, it is likely that our alignment follows magnetic structures reasonably well, so that the spatial distribution of chromospheric heating can be related to the spatial distribution photospheric magnetic flux.
\section{Results} \label{sec:result}

In order to distinguish the heating mechanisms operating in the chromosphere over plage regions, we investigate relations between the heating energy flux and the magnetic field in the photosphere and the chromosphere.
Figure \ref{fig.result1} shows maps of the total integrated radiative cooling energy flux of the \ion{Mg}{2} h \& k lines, the chromospheric magnetic field strength, $B_{{\rm ch}}$, the photospheric magnetic field strength, $B_{{\rm ph}}$, the photospheric line-of-sight velocity, the solar vertical component of the electric current density in the photosphere, $J_{Z, {\rm ph}}$, and the horizontal component of the electric current density, $J_{H}$.
For the velocity, blue and red indicate velocities of the \ion{Si}{1} moving toward and away from the observer, respectively. 
Since the thermal conduction is negligible in the chromosphere, the heating energy flux should balance with the radiative cooling energy flux in a steady state \citep{priest82}.
The \ion{Mg}{2} h \& k radiative cooling energy flux maps display where chromospheric heating occurs because about 15\% of the radiative energy is released as emissions of the \ion{Mg}{2} lines \citep{avrett81, vernazza81}.
The rest of the radiative energy is mainly released as \ion{H}{1} and \ion{Ca}{2} lines.
In the photospheric maps (the three bottom rows), the red contour lines indicate the place where the chromospheric heating energy flux is large when shifted by the displacement derived in section \ref{sec:method:align} to compare the \ion{Mg}{2} flux with photospheric structures.

From left to right, the columns of Figure \ref{fig.result1} display the mapped regions for Plage A at 9:27 UT, Plage B at 9:30 UT, Plage B at 9:34 UT, and Plage C, i.e., the periphery of the plage region, at 9:38 UT.
In Plage B, the magnetic field structure in the photosphere and the chromosphere, and both components of the current density structure were stable, while the area with large heating energy flux changed in 4 minutes.

\begin{figure}
\begin{center}
\includegraphics[angle=0,scale=1.0,width=80mm]{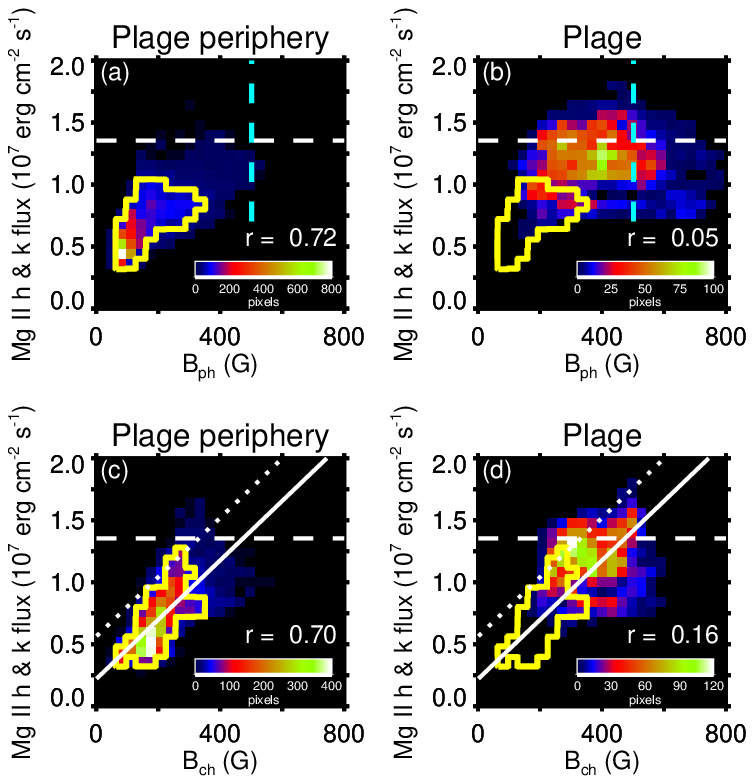}		
\end{center}
\caption{
Density plots of the total integrated radiative cooling energy flux of the \ion{Mg}{2} h \& k vs. (top) the magnetic field strength in the photosphere and (bottom) in the chromosphere.
Left and right columns show results for the periphery and the plage region, respectively.
The white solid lines show a linear fit to the relation between the \ion{Mg}{2} flux and $B_{{\rm ch}}$ in the periphery of the plage region, and the white dotted lines mark an upper bound of the \ion{Mg}{2} flux that deviates from the fit by less than twice of the standard deviation of the residual.
The white dashed horizontal and cyan dashed vertical lines indicate the \ion{Mg}{2} h \& k flux of $1.35 \times 10^{7} {\rm \,erg\,cm^{-2}\,s^{-1}}$ and $B_{{\rm ph}}$ of $500$ G drawn as red and blue contours in Figures \ref{fig.result1}, respectively.
The yellow contour lines are drawn at a density of 50 pixels with reference to the maps for the plage periphery.
		}
\label{fig.result2}
\end{figure}
\begin{figure*}
\begin{center}
\includegraphics[angle=0,scale=1.0,width=140mm]{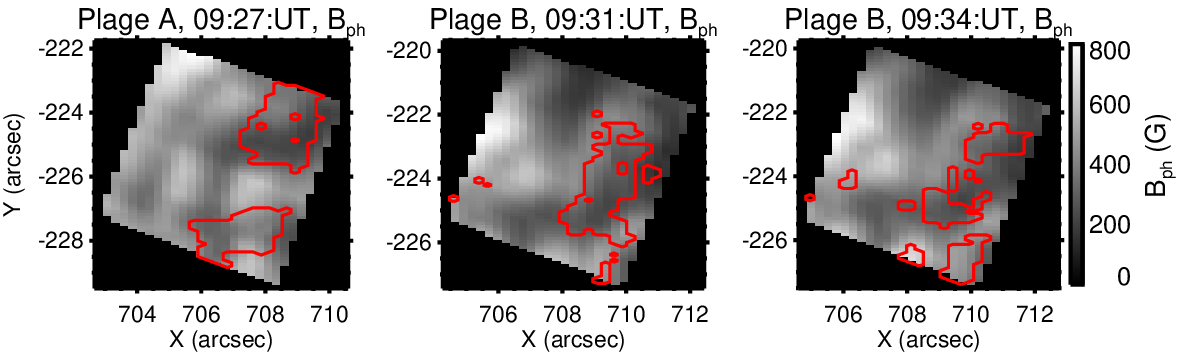}		
\end{center}
\caption{
Red contours surround regions where the \ion{Mg}{2} h \& k flux was larger than the white dotted line in Figure \ref{fig.result2} d.
The background maps are for the magnetic field strength in the photosphere within Plage A or B.
		}
\label{fig.result3}
\end{figure*}

\subsection{Chromospheric radiative flux vs. magnetic field strengths}

We compare chromospheric heating to the photospheric and chromospheric field strengths for both the plage interior and plage periphery data sets.
We find that the plage periphery exhibits a roughly linear trend between heating and magnetic field strength, while the trend breaks down when for the plage-interior regions.
As we argue below, this change in trend is not due to noise, and therefore may reflect a true physical relationship.

The photospheric magnetic field is strong in down flow regions (red in the Figure \ref{fig.result1} maps $V_{{\rm ph}}$) where granule-scale convection gathers the magnetic flux to the inter-granular lanes and forms strong magnetic field concentrations.  This effect is more pronounced in the plage interior (Plage A and B) compared to the periphery (Plage C).

Visual comparison between the \ion{Mg}{2} heating at the magnetic field gives somewhat mixed results. Clearly the strongest heating is often not co-spatial with the strongest magnetic field strength: compare the red contours of \ion{Mg}{2} flux to the field strengths in Figure \ref{fig.result1} rows 2 and 3.
In the plage interior there appears to be a preference for strong \ion{Mg}{2} h \& k flux at the edges or between patches of strong photospheric magnetic field (the blue contours in Figure \ref{fig.result1} row 3), but this was not always the case.  The spatial correspondence between the \ion{Mg}{2} h \& k flux and $B_{{\rm ch}}$ is less conclusive than that to $B_{{\rm ph}}$, due to the more uniform field strengths found in the chromosphere.

To better tease out real correlations we generated 2D histograms of the \ion{Mg}{2} flux versus $B_{{\rm ph}}$ and $B_{{\rm ch}}$ in the plage interior and periphery.
Figure \ref{fig.result2} shows the results, where the color scale indicates the number of pixels in each bin out of 4150 and 8300 total pixels for the plage and periphery regions, respectively.
The histograms involving $B_{\rm ph}$ include the displacement derived in section \ref{sec:method:align}.
As reflected in the figure, we recover the general trend supporting a non-linear increase in chromospheric energy flux versus the photospheric field strength, as reported by \citet{skumanich75}, \citet{schrijver89}, and \citet{barczynski18}.
Here that trend is also reflected in the trend between the radiated flux and $B_{\rm ch}$.
We note that in some cases, the magnetic field in the chromosphere is stronger than in the photosphere, i.e. at a given spatial pixel. 
This is particularly the case in the periphery, and we expect it can be explained by the horizontal expansion of the magnetic flux tubes in the chromosphere.

In the periphery of the plage region, the \ion{Mg}{2} h \& k flux had strong positive correlation with $B_{{\rm ph}}$ and $B_{{\rm ch}}$ with linear Pearson correlation coefficients, $r$, of 0.72 and 0.70, respectively.
We fitted the relation between the \ion{Mg}{2} flux and $B_{\rm ch}$ with a linear function with a proportionality factor equal to $(2.38 \pm 0.03) \times 10^4 {\rm \, erg \, cm^{-2} \, s^{-1} \, G^{-1} }$ (the white solid line in Figure \ref{fig.result2}c).
The dotted white line shows the same relation but shifted in $+Y$ direction from the linear fit by two times of the standard deviation of the residual.
The white solid and dotted lines in Figure \ref{fig.result2}d are the same as those in Figure \ref{fig.result2}c.

The \ion{Mg}{2} h \& k flux and the magnetic field strengths are continuously distributed from the periphery to the plage on the density plot (see yellow contour lines in panels b and d), since a part of the field-of-view for Plage B overlapped that for Plage C (Figure \ref{fig.obs_ar}).  
In the plage region, the magnetic field strengths and the \ion{Mg}{2} h \& k flux were stronger than those in the periphery of the plage region, and they did not have significant correlation in contrast to the periphery (Figure \ref{fig.result2} b and d).
Since the errors of $B_{{\rm ph}}$ and $B_{{\rm ch}}$ are $30$ - $40$ G at $B=400$ G (Figure \ref{fig.cal_error}a) and the error of \ion{Mg}{2} h \& k flux is $0.6$ \% (Section \ref{sec:method:mg2}), the lack of correlation cannot be explained by the observational noise on these spatial scales; rather, it may be reflective of the true physical relationship.
Moreover, the high density peak above the 2-sigma fit in Figure \ref{fig.result2}d (dotted line) shows that the \ion{Mg}{2} h \& k flux in the plage interior is significantly stronger than the linear relation derived from the plage periphery region.

To isolate the regions with significantly enhanced radiative flux we selected regions that are above the dotted line in Figure \ref{fig.result2}d.
In Figure \ref{fig.result3} we overplot contours of these high-flux regions on the maps of $B_{{\rm ph}}$ (after accounting for spatial displacements described in section \ref{sec:method:align}).
This figure demonstrates that the strongest heating within the plage interior regions appears to come from either the edges or in-between the strong photospheric concentrations.

\begin{figure}
\begin{center}
\includegraphics[angle=0,scale=1.0,width=80mm]{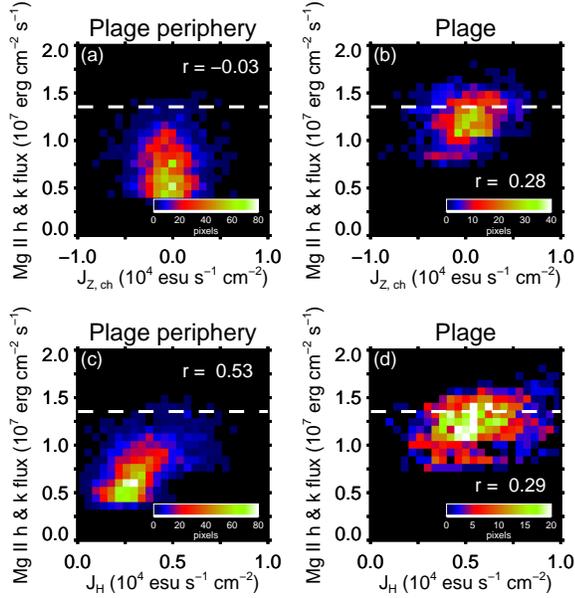}		\end{center}
\caption{
Density plots of the total integrated radiative cooling energy flux of the \ion{Mg}{2} h \& k vs. the vertical component of the electric current density in the chromosphere (top) and the horizontal component of the electric current density (bottom).
Left and right columns show results for the periphery and the plage region, respectively.
The white dashed horizontal lines indicate the \ion{Mg}{2} h \& k flux of $1.35 \times 10^{7} {\rm \,erg\,cm^{-2}\,s^{-1}}$ drawn as red contours in Figures \ref{fig.result1}.
		}
\label{fig.result4}
\end{figure}

\subsection{Chromospheric radiative flux vs. electric currents}

In regions where the \ion{Mg}{2} h \& k flux was large, $J_{Z, {\rm ph}}$ was weak and had positive or negative signs, i.e., was consistent with zero, while the strong photospheric patches correlate relatively well with strong positive $J_{Z, {\rm ph}}$ (Figure \ref{fig.result1}).
In particular, strong \ion{Mg}{2} h \& k flux seems to be emitted from areas around strong $J_{Z, {\rm ph}}$ located at ($707"$, $-224"$) with changing the shape of the strong \ion{Mg}{2} h \& k flux region (see the three left columns in the bottom row of Figure \ref{fig.result1}).
The horizontal component of the electric current density, $J_{H}$, was strong around photospheric magnetic patches.
However, the region of the strong $J_{H}$ does not match the region where the \ion{Mg}{2} h \& k flux was strong. 

Figure \ref{fig.result4} shows two-dimensional histograms of the \ion{Mg}{2} h \& k flux versus $J_{H}$ and the vertical component of the electric current density in the chromosphere, $J_{Z, {\rm ch}}$ for the plage region and its periphery.
The \ion{Mg}{2} h \& k flux and $J_{H}$ in the periphery of the plage region showed a more modest correlation ($r=0.53$) than between the \ion{Mg}{2} h \& k flux and the $B_{ch}$ ($r=0.70$, see Figure \ref{fig.result2}).
Otherwise, there was no significant correlation.
Since the $1 \sigma$ error of  $J_{{\rm Z, ch}}$ for the periphery and the plage regions are $2 \times 10^3$ and $3 \times 10^3$ esu s$^{-1}$ cm$^{-2}$, respectively (Figure \ref{fig.pdf_error_current}), we might not be able to find significant relations in panels (a) and (b) of Figure \ref{fig.result4} due to sensitivity of the measurements.

\subsection{Temporal variations of the magnetic field and the line-of-sight velocity}

\begin{figure}
\begin{center}
\includegraphics[angle=0,scale=1.0,width=80mm]{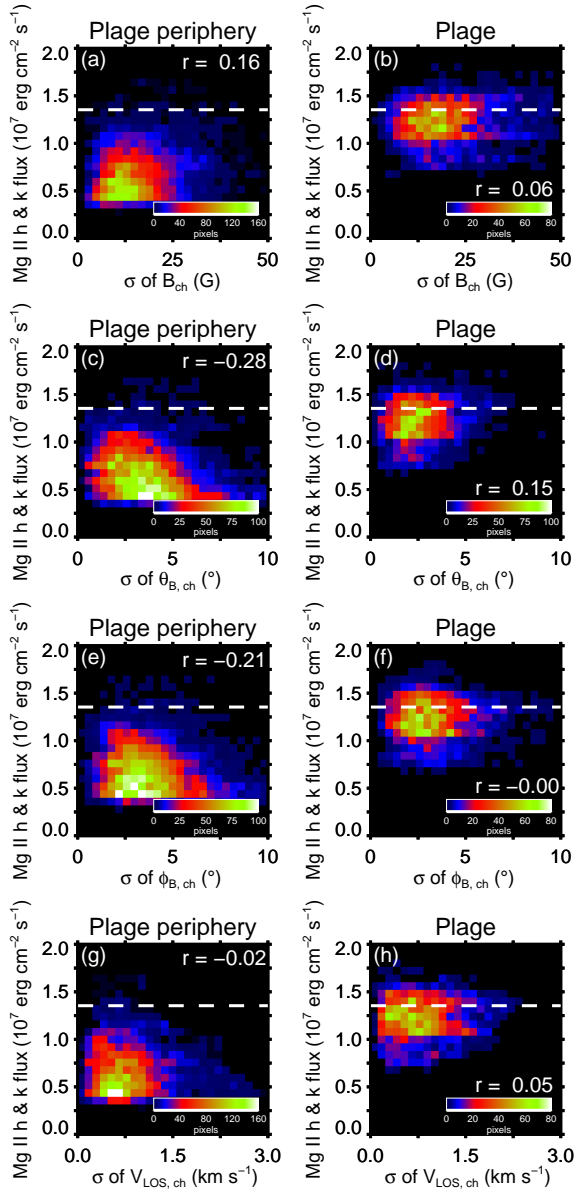}
\end{center}
\caption{
Density plots of the total integrated radiative cooling energy flux of the \ion{Mg}{2} h \& k vs. the standard deviation of the magnetic field strength, the inclination and the azimuth angles of the magnetic field in the solar vertical reference frame, and the line-of-sight velocity in the chromosphere (from top to bottom).
Left and right columns show results for the periphery and the plage region, respectively.
}
\label{fig.result5}
\end{figure}

We investigate not only spatial relations between the \ion{Mg}{2} flux and the magnetic field structures but also relations of the \ion{Mg}{2} flux to amplitudes of temporal variations in the chromospheric magnetic field and the velocity.
If there is a significant correlation among them, waves can be at work. 
Figure \ref{fig.result5} shows two-dimensional histograms of the \ion{Mg}{2} h \& k flux versus the standard deviation of temporal variation of physical quantities in $104$ seconds before IRIS scanned the region.
We do not find any significant correlation between the \ion{Mg}{2} h \& k flux and the standard deviations of the magnetic field or the line-of-sight component of the velocity in the chromosphere.

\section{Summary \& Discussion} \label{sec:discussion}

Basal heating of chromospheric plage regions requires the largest quasi-steady heating rate of any region of the solar atmosphere from the photosphere through the corona. 
Table \ref{table.mechanism1} lists many proposed plage heating mechanisms and how heating relates to the magnetic field.
High spatial resolution observations of chromospheric magnetic fields and heating are crucial for distinguishing between the proposed mechanisms.
We inferred the magnetic field vector in the photosphere and the chromosphere over a plage region using spectro-polarimeteric data taken with the integral field unit mode of GRIS installed at the GREGOR telescope. The integrated radiative cooling energy flux of the \ion{Mg}{2} h \& k emissions, which is used as a proxy for the total radiative losses that balance the deposited energy, was calculated from the spectra obtained by IRIS.

\begin{figure}
\begin{center}
\includegraphics[angle=0,scale=1.0,width=80mm]{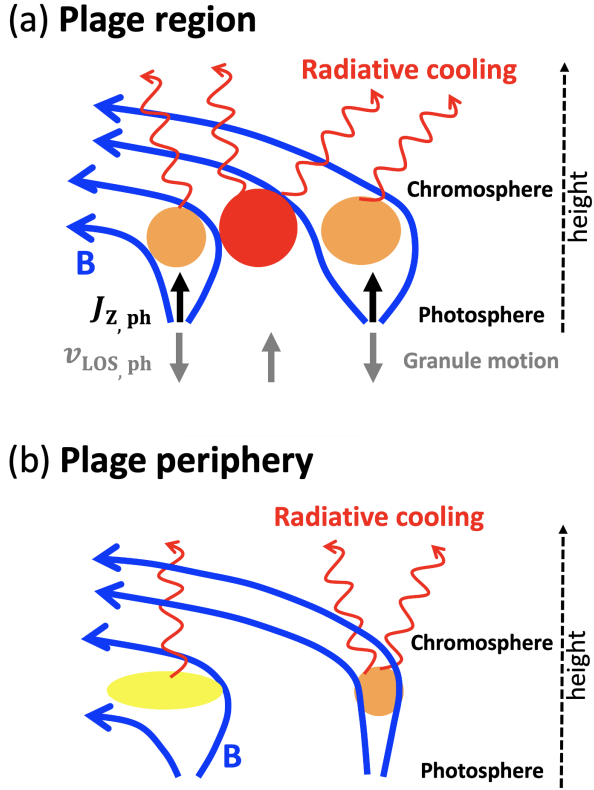}
\end{center}
\caption{
Relation between the radiative cooling in the chromosphere and the magnetic structures in the photosphere and the chromosphere (a) in the plage region and (b) in the plage periphery. 
The thick blue lines labeled B signify magnetic field lines. 
The thick black and gray arrows indicate $J_{Z, {\rm ph}}$ and $V_{{\rm LOS, ph}}$, respectively. 
The colored ovals indicate heated chromosphere.
The redder the color in the ovals, the stronger the heating.
		}
\label{fig.cartoon1}
\end{figure}

Figure \ref{fig.cartoon1} provides a schematic summarizing our findings. In the periphery of the plage region within the limited field of view seen by GRIS we found that the radiative cooling energy flux strongly correlates with $B_{{\rm ph}}$ and $B_{{\rm ch}}$ with linear Pearson correlation coefficients of 0.72 and 0.70, respectively. The linear relation holds over an extensive range of chromospheric field strengths (50 - 400 G) with a proportionality factor of $(2.38 \pm 0.03) \times 10^4 {\rm \, erg \, cm^{-2} \, s^{-1} \, G^{-1} }$.
We found somewhat different behavior in the plage interior regions (Plage A and B):
\begin{enumerate}
\item The radiative cooling energy flux, i.e, heating energy flux, did not have significant correlation with $B_{{\rm ch}}$.
\item The heating energy flux was typically larger between magnetic patches in the photosphere or at the edges of the patches, but this was not always the case.
\item Between the magnetic patches or at their edges, the heating energy flux is significantly larger than expected from the linear function of $B_{{\rm ch}}$, which is derived in the plage periphery region.
\item Strong heating seems to occur around a magnetic patch that has strong electric current density in the photosphere.
\item Regions where strong heating occurs were observed to change their shapes on timescales of approximately 4 minutes.
\item The heating energy flux and the magnetic field strengths were stronger than those in the periphery of the plage region.
\item The heating energy flux did not have significant correlation with $J_{H}$ or $J_{Z, {\rm ch}}$. 
\item The heating energy flux also did not correlate with the standard deviation over temporal variations of physical quantities in the chromosphere.
\item We verified that $B_{{\rm ph}}$ is strong in the down flow regions, which are inter-granular regions. We also verified that $B_{{\rm ch}}$ was stronger than $B_{{\rm ph}}$ in some cases.
\end{enumerate}

Based on these findings, we discuss what mechanisms may heat the chromosphere in the plage region and its periphery. For the periphery of the plage region, the heating energy positively correlates with $B_{{\rm ch}}$ (Figure \ref{fig.result2}).
Positive correlation between $B_{{\rm ph}}$ and the radiation from \ion{Ca}{2} H \& K have been also reported at network regions where the magnetic field has an important role for heating in the chromosphere \citep[e.g., ][]{rezaei07, beck13}.
Assuming the magnetic field in the corona is the same as that in the upper chromosphere, magnetic flux tubes in the plage periphery can be heated by Alfv\'en wave turbulence or by collisions between ions and neutral atoms relating to propagating torsional Alfv\'en waves \citep[Table \ref{table.mechanism1}, ][]{ballegooijen11, soler17}.
Interestingly, both mechanisms relate to Alfv\'en waves driven at or below the photosphere.
Although magnetic reconnection could also be responsible for the heating, it would have to be in the strong-guide field regime \citep{katsukawa07penumbra, sakai08, nishizuka12}: the fraction of horizontal or opposite polarity field is too small in our observations to produce large-scale anti-parallel magnetic fields (Figure \ref{fig.thb_histogram}).
It is still possible that small and weak opposite polarity patches were hidden beneath the expanding flux tubes \citep{buehler15, wang19, chitta19}.

The two-dimensional histogram of the \ion{Mg}{2} h \& k radiative energy flux versus $B_{{\rm ch}}$ for the plage region lies approximately along the linear trend derived from the radiative energy flux and $B_{{\rm ch}}$ in the periphery region (panels c and d of Figure \ref{fig.result2}).
These findings suggest that the same mechanisms can heat both the plage interior and periphery regions.
However, we find no significant correlation between the radiative energy flux and $B_{{\rm ch}}$ within the plage region (finding 1).
In addition, radiative energy flux that was significantly larger than expected from the periphery region was emitted from between magnetic patches or at edges of the magnetic patches in the photosphere (findings 2 and 3).
This suggests that other additional mechanisms can heat the plage chromosphere between the patches or at their edges.

Based on the heating location near or between flux tubes, possible heating mechanisms include shock wave heating due to interactions between magnetic flux tubes \citep{snow18}, ion-neutral collisions at vortices produced by resonant absorption or Kelvin-Helmholtz instability \citep{kuridze16}, or the ion-neutral collisions related to electric current frequently located along edges of flux tubes \citep{khomenko18} (see Table \ref{table.mechanism1}).
We propose the following mechanism in this paper putting together these three processes.
The observed magnetic flux tubes can be perturbed or twisted by the granule motions.
Such perturbations will propagate upward along the tubes \citep[e.g., ][]{shelyag16} and produce vortices in the chromosphere \citep{yadav21}.
Resonant absorption or Kelvin-Helmholtz instability might also contribute to the vortex formation at the boundaries of the tubes \citep[e.g., ][]{kuridze16}.
When multiple flux tubes interact or merge at the chromosphere, velocity and magnetic field structures of the vortices present in each tube become superposed and generate complex substructures between flux tubes \citep{snow18}.
Such substructures would accompany electric current density perpendicular to the magnetic field, where the magnetic energy strongly dissipates due to the ion and neutral collisions \citep{khomenko18, yadav21}.

The above scenario is supported by the lack of a significant correlation between the heating energy flux and $B_{{\rm ch}}$ in the plage region (finding 1).
This is consistent with what \citet{yadav21} found in their simulation for chromospheric heating due to electric currents associated with vortex interactions, although they studied vortex interactions inside a magnetic flux tube.

\citet{snow18} proposed a chromospheric heating mechanism due to upward propagating shock waves with a flow speed of $\sim 50$ km s$^{-1}$, which is generated by superposition of vortices between magnetic flux tubes.
However, no such high speed upward movement in the plage chromosphere was observed by GRIS for 104 seconds before IRIS scanned the region. Because the high-speed plasma can pass the formation layer of the \ion{He}{1} line in a short time, our time cadence of $26$ seconds might be too long to detect the signature of the shock wave.

Since our observation showed that the heating changed in time, and that the heating energy was not always large in all regions between the magnetic patches and at their edges (findings 2 and 5), determining the location where the additional heating occurs is a necessary but not sufficient condition for distinguishing between different heating mechanism.
In the plage, oscillations in intensity of spectral lines formed in the chromosphere occur with a period of $\sim 5$ minutes \citep{bhatnagar72}.
\citet{kostik16} reported that the brightness of \ion{Ca}{2} H correlates with the power of 5 minute oscillation in the photosphere.
As \citet{tarr17} and \citet{tarr19} described in detail, magneto acoustic energy can dissipate at topological features of the magnetic field, e.g., magnetic nulls, magnetic minima, or separatrix surfaces. 
This allows another pathway to spatially and temporally localize heating that would act in combination with the mechanisms above but modify them based on the global magnetic connectivity.

One requirements for the formation of a plage region is a closed field region that connects opposite polarity patches through a hot dense corona \citep{carlsson19}.
Indeed, it is implied by our observation that the plage region was connected to the opposite polarity region (section \ref{sec:method:b:inversion}).
It is possible that some aspects of the overlying corona may play an important role in heating of the plage regions.

We found that all strong photospheric flux regions have the same sign vertical photospheric current  density (Figure \ref{fig.result1} and \ref{fig.cartoon1}).
The expansion of same-twist photospheric flux tubes will result in anti-parallel (plus guide-field) magnetic fields at the boundaries of the flux domains in the chromosphere.
This can lead to magnetic energy dissipation due to ion-neutral collisions or magnetic reconnection \citep[e.g., ][]{leake13, ni20}.


We do not find a correlation between the radiative energy flux and the electric current density (finding 7, Figure \ref{fig.result1} and \ref{fig.result4}).
Because the width of an electric current sheet is only a few 10s of kms and their polarity is mixed at interfaces of the vortices (see Figure 6 of \citet{yadav21}), the spatial resolution of our observations is not enough to resolve the current sheets.
We also did not find perturbations in the velocity or the magnetic field on magnetic flux tubes, which is expected in our proposed process (Figure \ref{fig.result5}).
\citet{shelyag16} derived amplitudes for magnetic field fluctuations due to Alfv\'en and fast-mode magnetoacoustic waves up to 5 G, which is too small to be detected using the present measurements (Figure \ref{fig.cal_error}).
Upcoming observations with the National Science Foundation's Daniel K. Inouye Solar Telescope will enable us to search for and resolve electric current sheets or the magnetic field fluctuations \citep[DKIST, ][]{rimmele20}.

  

\acknowledgments
The authors thank Ms. K. Gerber for operating the GREGOR telescope during the observation.
We also thank Dr. X. Sun, Dr. M. Kramar, and Dr. S. Mahajan, who discussed with us every week. 
The authors are grateful to Dr. A. Asensio Ramos for his guidance in the use of HAZEL v2.0.
TA extends thanks to Dr. S. Nagata for discussing various science objectives during tea time at Hida observatory. The concept in this paper developed from those discussions.
The National Solar Observatory (NSO) is operated by the Association of Universities for Research in Astronomy, Inc. (AURA), under cooperative agreement with the National Science Foundation.
The GREGOR solar telescope used in this study was built by a German consortium under the leadership of the Leibniz-Institute for Solar Physics (KIS) in Freiburg with the Leibniz Institute for Astrophysics Potsdam, the Institute for Astrophysics G${\rm \ddot{o}}$ttingen, and the Max Planck Institute for Solar System Research in G${\rm \ddot{o}}$ttingen as partners, and with contributions by the Instituto de Astrof\'isica de Canarias and the Astronomical Institute of the Academy of Sciences of the Czech Republic. 
The SDO data are provided courtesy of NASA/SDO and the AIA and HMI science teams.
This work was supported by JSPS KAKENHI Grant Number 20KK0072.

%

\vspace{5mm}
\facilities{GREGOR(GRIS)}


\software{Solar Soft}

\bibliography{s2_2}



\end{document}